\documentclass[12pt]{article}
\usepackage[T2A]{fontenc}
\usepackage[utf8]{inputenc}
\usepackage[english]{babel}
\usepackage{amssymb}
\usepackage{amsmath}
\usepackage{graphicx}
\usepackage[small,sc,center]{titlesec}
\usepackage{indentfirst}
\usepackage{siunitx}
\usepackage[labelsep=period]{caption}
\usepackage{threeparttable}

\renewcommand{\thefootnote}{\fnsymbol{footnote}}

\newcommand{\nc}{\newcommand}
\nc{\ek}{E_\mathrm{K}}
\nc{\mzams}{M_\mathrm{ZAMS}}
\nc{\teff}{T_\mathrm{eff}}
\nc{\tev}{t_\mathrm{ev}}
\begin{document}

\begin{center}
\textbf{Secular perion changes and fundamental parameters of long--period Cepheids}

\vskip 3mm
\textbf{Yu. A. Fadeyev\footnote{E--mail: fadeyev@inasan.ru}}

\textit{Institute of Astronomy, Russian Academy of Sciences,
        Pyatnitskaya ul. 48, Moscow, 119017 Russia} \\

Received June 13, 2018
\end{center}

\textbf{Abstract} ---
Hydrodynamic computations of nonlinear Cepheid pulsation models with periods from
20 to 100 day on the evolutionary stage of core helium burning were carried out.
Equations of radiation hydrodynamics and time--dependent convection were solved with
initial conditions obtained from selected models of evolutionary sequences of population I
stars with initial masses from $8M_\odot$ to $12.5M_\odot$.
For each crossing of the instability strip the pulsation period $\Pi$ and the rate
of period change $\dot\Pi$ were derived as a function of evolutionary time.
Comparing results of our computations with observational estimates of $\Pi$ and $\dot\Pi$
we determined fundamental parameters (the age, the mass, the luminosity and the radius)
of seven long--period Cepheids.
Theoretical estimates of the stellar radius are shown to agree with radius measurements
by the Baade--Wesselink technique within 3\% for RS~Pup and GY~Sge whereas for SV~Vul
the disagreement between theory and observations does not exceed 10\%.


\newpage
\section{introduction}

Together with period--luminosity relation $\delta$~Cep pulsating variables (Cepheids)
obey another fundamental dependence relating the pulsation period and the stellar age.
Existence of the negative correlation between the stellar age and the pulsation period
was established due to the fact that some Cepheids were found to belong to young
clusters and associations (Hodge 1961; Efremov 1978).
These conclusions were confirmed by stellar evolution computations
(Kippenhahn, Smith 1969; Meyer--Hofmeister 1969; Bono et al. 2005),
so that the pulsation period of the Cepheid belonging to the cluster can be considered
as a reliable indicator of the cluster age.

Of greatest interest among $\delta$~Cep pulsating variables are long--period Cepheids
because many of them belong to young clusters and OB associations and at the same time
they can be observed at great distances due to their high luminosity.
Another important property of long--period Cepheids is the high rate of period change
which is an indicator of the rapid stellar evolution.
Bright Cepheids are regularly observed for more than a century, so that for some of them
reliable observational estimates of period change rates became available.
One of Cepheids with the highest period change rate ($\dot\Pi=719$~s/yr) is II~Car
pulsating with period $\Pi=64.4$ day (Berdnikov, Turner 2010).

Long--period Cepheids are distant objects, so that their fundamental parameters remain
uncertain due to significant errors in their trigonometric parallaxes.
In our previous works (Fadeyev 2015a; 2015b) we have shown that the stellar age
$\tev$, the mass $M$, the luminosity $L$ and the radius $R$ of the pulsating star
can be determined from consistent evolutionary and nonlinear stellar pulsation calculations
provided that the pulsation period and the rate of period change are obtained from
observations.
In the present work this method we employ to determine the fundamental parameters of
seven Cepheids with periods from 33 to 64 day.
Period change rates of these pulsating variables were evaluated from $O-C$ diagrams
based on archive data for more than the century 
(Berdnikov 1994; Turner, Berdnikov 2004; Berdnikov et al. 2007, 2009a, 2009b;
Berdnikov, Turner 2010).
All Cepheids considered in our study are on the evolutionary stage of core helium burning.
Following to common terminology we call them as Cepheids on the stage of the second
crossing ($\dot\Pi < 0$) and the third crossing ($\dot\Pi > 0$) of the instability strip.

\section{methods of computation}

Solution of the Cauchy problem for equations of radiation hydrodynamics and
time--dependent convection for radial stellar oscillations was carried out with
initial conditions obtained from stellar evolution calculations.
To this end we computed 20 evolutionary tracks from the main sequence to central
helium exhaustion for stars with initial masses $8M_\odot\le\mzams\le 12.5M_\odot$
and initial composition $X=0.7$, $Z=0.02$ where $X$ and $Z$ are relative mass
fractions of hydrogen and elements heavier than helium.

Evolutionary computations were done with the MESA code version 10398 (Paxton et al. 2018).
We have followed the nucleosynthesis of 29 isotopes from hydrogen ${}^1\mathrm{H}$
to aluminium ${}^{27}\mathrm{Al}$ with 51 reactions.
The reaction rates were taken from the JINA Reaclib database (Cyburt et al. 2010).
Convective mixing was treated through the mixing length theory (B\"ohm--Vitense 1958)
with a mixing length to pressure scale height ratio $\alpha_\mathrm{MLT}=1.8$.
Additional mixing on the convection stability boundaries due to overshooting
was taken into account using the prescription of Herwig (2000) with
exponential parameter $f=0.016$.
The mass loss rate was calculated following the prescription of Reimers (1975)
with mass loss parameter $\eta_\mathrm{R}=0.5$.

In the present study we considered self--exciting radial stellar oscillations
arising in the case of pulsational instability due to small initial hydrodynamic
perturbations.
Each solution of the equations of hydrodynamics is carried out with initial conditions
represented by a selected model with age $\tev$ of the computed evolutionary sequence.
Evolutionary computations were carried out with the number of mass zones
$N\sim 5\times 10^3$ which in general varies each step following the changes of
the stellar structure.
Equations of hydrodynamics were solved on the Lagrangian grid with the fixed number
of the nodes ($N=400$).
It should be noted that in evolutionary computations the stellar model is considered
from the center to the surface whereas equations of hydrodynamics for radial oscillations
are solved only for the outer layers of the star because the amplitude of radial
displacement exponentially decreases inwards.
In this work the inner boundary of the hydrodynamic model was set in the layer with
relative radius $r/R=0.1$ where $R$ is the radius of the outer boundary of the
evolutionary model.

Transition from the evolutionary model to the initial hydrodynamic model was done
using nonlinear interpolation of basic variables (the radius, the pressure, the temperature,
the gas density etc.) with respect to mass coordinate and where interpolation errors played
the role of initial perturbations.
The total kinetic energy $\ek$ of initial gas flows was less than 1\% of the maximum
kinetic energy during limit cycle oscillations.

In the end of calculation of the hydrodynamic model we determined the pulsation period
$\Pi$ using the discrete Fourier transform of the kinetic energy on the time interval
$10^2\lesssim t/\Pi\lesssim 10^3$.
Together with period we evaluated the instability growth rate $\eta=\Pi d\ln\ek/dt$.
Inverse $\eta^{-1}$ is equal to the $e$--folding time of the kinetic energy
expressed in units the pulsation period $\Pi$.
In the case of pulsational instability the growth rate $\eta$ was evaluated for the
time interval with linear growth of $\ln\ek$.
For decaying oscillations ($\eta<0$) the maxima of $\ln\ek$ decrease with time
nearly linearly within the whole integration interval.

The system of the equations of hydrodynamics and parameter selection for the
transport equations of the time--dependent convection model (Kuhfu\ss 1986)
are discussed in our previous papers (Fadeyev 2013; 2015a).

\section{results of computations}

The principal goal of our evolutionary and hydrodynamic computations is that
to determine the period of radial oscillations as a function of evolutionary time
$\tev$ within the boundaries of the instability strip.
For each crossing of the instability strip we selected nearly 10 models of the
evolutionary sequence and solved the equations of hydrodynamics for each of them.
Boundaries of the instability strip, that is the evolutionary time $\tev$ corresponding
to the growth rate $\eta=0$  were determined by interpolation between two adjacent
models with opposite signs of $\eta$.
Dependence of the pulsation period on the evolutionary time was fitted to a good
accuracy by polynomials of the second order using the least square method.
In general the procedure is almost the same as that employed in our earlier study
devoted to less massive Cepheids of the Large Magellanic Cloud (Fadeyev 2013) and
the only difference is that the long--period Cepheids remain the fundamental mode
pulsators while they cross the instability strip.

\begin{figure}
\centerline{\includegraphics[width=15cm]{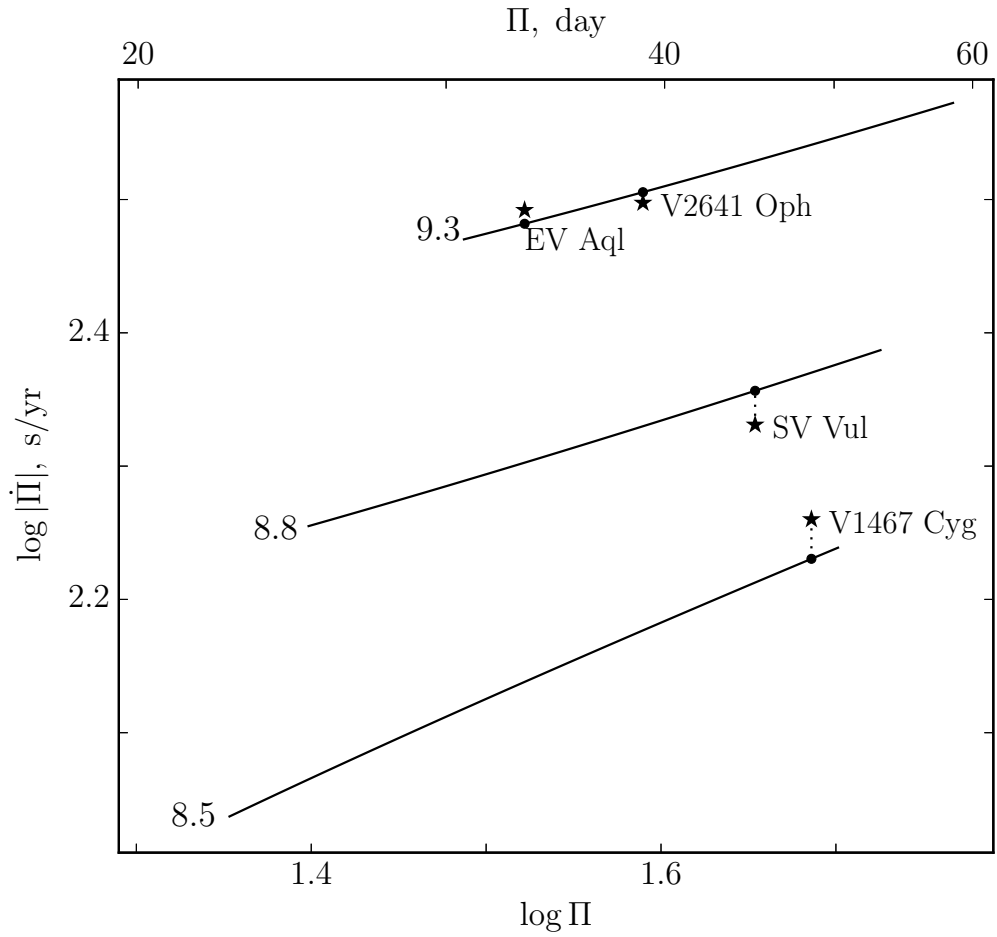}}
\caption{The rate of period change against the pulsation period for
         Cepheid models during the second crossing of the instability strip.
         Initial masses $\mzams$ are indicated near the curves.}
\label{fig1}
\end{figure}

\begin{figure}
\centerline{\includegraphics[width=15cm]{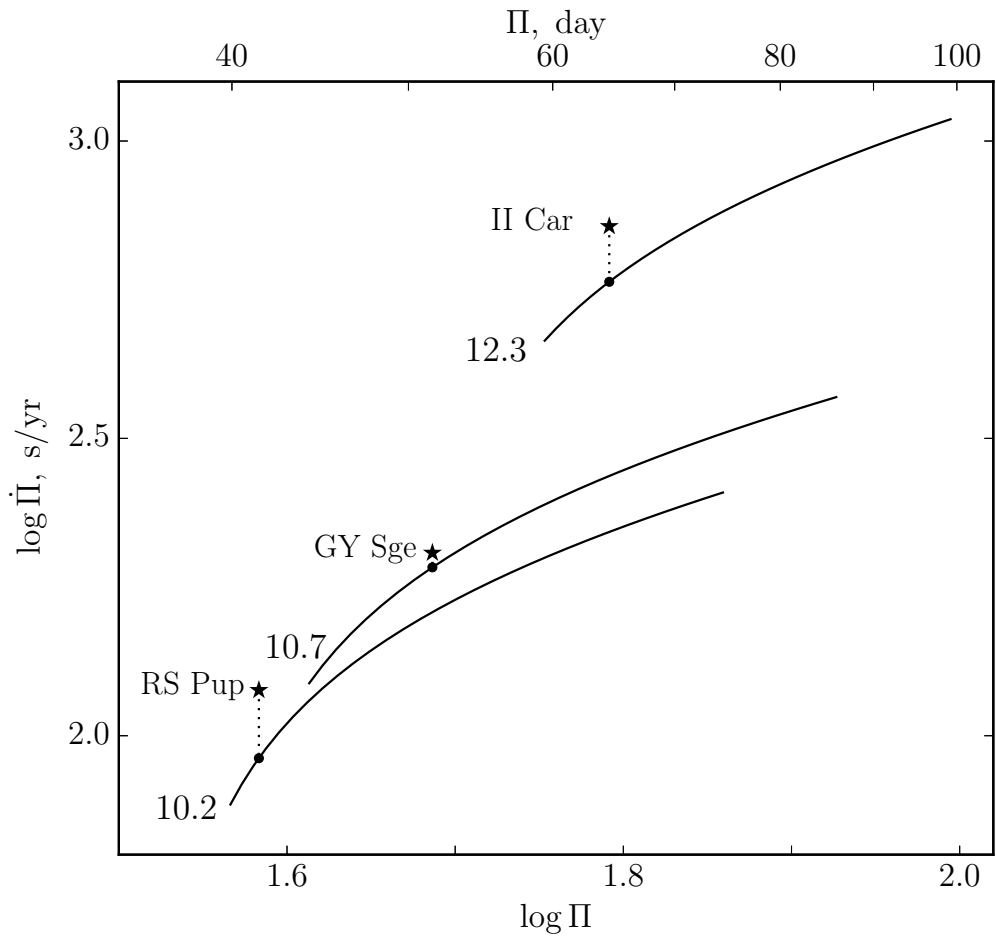}}
\caption{Same as Fig.~\ref{fig1} but for Cepheid models during the third crossing
         of the instability strip.}
\label{fig2}
\end{figure}

Results of our calculations are illustrated in Figs.~\ref{fig1} and \ref{fig2} where
the rate of period change is shown as a function of pulsation period for the second
and the third crossings of the instability strip, respectively.
Each curve of these plots represents the change of the rate of period change $\dot\Pi$ as
a function of period $\Pi$ while the star evolves between the instability strip edges.
During the second crossing of the instability strip (Fig.~\ref{fig1}) stellar evolution
is accompanied by decrease of $\Pi$, so that along the vertical axis we give
the logarithm of the absolute value of the period change rate.
During the third crossing of the instability strip (Fig.~\ref{fig2}) the pulsation period
increases as the star evolves.
Typical effective temperatures of hydrodynamic models range within
$4000\:\mathrm{K}\lesssim\teff\lesssim 5300\:\mathrm{K}$.

To compare results of our calculations with observations we indicate in Figs.~\ref{fig1}
and \ref{fig2} observational estimates of the period and the period change rate of
long--period Cepheids.
Four of these stars (V1467~Cyg, SV~Vul, V2641~Oph, EV~Aql) are on the second crossing
of the instability strip and three (RS~Pup, GY~Sge, II~Car) are on the third crossing.
It should be noted that for the sake of graphical representation we show the tracks
best fitted to observations.
Filled circles on the tracks indicate the models of evolutionary sequences with
least deviation in $\dot\Pi$.
Basic parameters of these models are listed in the table where $\tev$ is the star age,
$M$, $L$ and $R$ are the mass, the luminosity and the radius,
$\dot\Pi$ и $\dot\Pi_\star$ are the theoretical and observational estimates of the
period change rate.
Comments to the table give references to works with observational estimates $\dot\Pi_\star$.

\section{conclusions}

Comparison of theoretical tracks on the diagram period--period change rate
with observations allowed us to determine the fundamental parameters of seven
long--period Cepheids.
However it should be noted that the grid of evolutionary tracks was computed
with step in initial mass of $\Delta\mzams=0.2M_\odot$ or $\Delta\mzams=0.3M_\odot$
therefore one has to evaluate the role of $\Delta\mzams$ in
uncertanty of our estimates.
For such a test the most appropriate is the Baade--Wesselink method providing
observational estimates of the stellar radius based on simultaneous spectroscopic
and photometric observations.
Such measurements were carried out for three Cepheids that were
considered in the present study.

The mean radius of the Cepheid RS~Pup measured by the Baade--Wesselink method
is $R=191R_\odot$ (Kervella et al. 2017) and exceeds the radius given in the table
by 2\%
The Cepheid RS~Pup is of special interest due to the fact that it is surrounded
by a diffuse reflection nebula (Westerlund 1961).
Measurements of the light echoes give the distance to RS~Pup of $d=1910$ pc with
relative accuracy of 4\% (Kervella et al. 2014).
The mean bolometric luminosity of RS~Pup is $L=2.17\times 10^4L_\odot$ (Kervella et al. 2017)
which is by 11\% higher than the model luminosity given in the table.
Bearing in mind uncertainties in the bolometric correction we can conclude that
the theoretical and observational estimates of the luminosity are in a good agreement.

At the same time we have to note the existing contradiction between the ages of the Cepheid
and the association Pup~OB1 so that membership of RS~Pup in the association
becomes doubtful.
Havlen (1972) has shown that the association has an age $\approx 4\times 10^6$ yr
whereas the age of RS~Pup determined by the period--age relation (Bono et al. 2005)
is several times larger and is $\approx 1.7\times 10^7$.
So large age difference obviously excludes any connection between RS~Pup and
the stellar association (Kervella et al. 2008).
As seen from the table the age of RS~Pup estimated in the present study 
($\tev\approx2.4\times 10^7$ лет) corroborates this conclusion.

Radius measurements for GY~Sge were done in two works where the authors obtained
almost the same values:
$R=210R_\odot$ (Sachkov 2002) and $R=208R_\odot$ (Rastorguev, Dambis 2011).
These observational estimates agree with our theoretical estimate of the radius
within 3\% and allow us to conlude a good agreement between the theory and
observations.

The mean radius of SV~Vul derived from a modified Baade--Wesselink technique is
$R=201R_\odot$ (Turner, Burke 2002) and exceeds the theoretical estimate of the
radius by less than 10\%.

\newpage
\section*{references}

\begin{enumerate}

\item L.N. Berdnikov, Astron. Lett. \textbf{20}, 232 (1994).

\item L.N. Berdnikov, E.N. Pastukhova, N.A. Gorynya, A.V. Zharova and D.G. Turner,
      Publ. Astron. Soc. Pacific \textbf{119}, 82 (2007).

\item L.N. Berdnikov, E.N. Pastukhova, D.G. Turner and D.J. Majaess,
      Astron. Lett. \textbf{35}, 175 (2009a).

\item L.N. Berdnikov, A.A. Henden, D.G. Turner and E.N. Pastukhova,
      Astron. Lett. \textbf{35}, 406 (2009b).

\item L.N. Berdnikov and D.G. Turner, Astron. Rep \textbf{54}, 392 (2010).

\item E. B\"ohm--Vitense, Zeitschrift f\"ur Astrophys. \textbf{46}, 108 (1958).

\item G. Bono, M. Marconi, S. Cassisi, F. Caputo, W. Gieren and G. Pietrzynski,
      Astrophys. J. \textbf{621}, 966 (2005).

\item R.H. Cyburt, A.M. Amthor, R. Ferguson, Z. Meisel, K. Smith, S. Warren, A. Heger,
      R.D. Hoffman, et al., Astrophys. J. Suppl. Ser. \textbf{189}, 240 (2010).

\item Yu.N. Efremov, Sov. Astron. \textbf{22}, 161 (1978).

\item Yu.A. Fadeyev, Astron. Lett. \textbf{39}, 746 (2013).

\item Yu.A. Fadeyev, MNRAS \textbf{449}, 1011 (2015а).

\item Yu.A. Fadeyev, Astron. Lett. \textbf{41}, 640 (2015b).

\item R.J. Havlen, Astron. Astrophys. \textbf{17}, 413 (1972).

\item F. Herwig, Astron. Astrophys. \textbf{360}, 952 (2000).

\item P. Hodge, Astrophys. J \textbf{133}, 413 (1961).

\item P. Kervella, A. M\'erand, L. Szabados, P. Fouqu\'e, D. Bersier, E. Pompei and G. Perrin,
      Astron. Astrophys. \textbf{480}, 167 (2008).

\item P. Kervella, H.E. Bond, M. Cracraft, L. Szabados, J. Breitfelder, A. M\'erand,
         W.B. Sparks, A. Gallenne, et al., Astron. Astrophys. \textbf{572}, A7 (2014).

\item P. Kervella, B. Trahin, H.E. Bond, A. Gallenne, L. Szabados, A. M\'erand,
      J. Breitfelder, J. Dailloux, et al., Astron. Astrophys. \textbf{600}, A127 (2017).

\item R. Kippenhahn and L. Smith, Astron. Astrophys. \textbf{1}, 142 (1969).

\item R. Kuhfu\ss, Astron. Astrophys. \textbf{160}, 116 (1986).

\item E. Meyer--Hofmeister, Astron. Astrophys. \textbf{2}, 143 (1969).

\item B. Paxton, J. Schwab, E.B. Bauer, L. Bildsten, S. Blinnikov, P. Duffell, R. Farmer,
      J.A. Goldberg, et al., Astrophys. J. Suppl. Ser. \textbf{234}, 34 (2018).

\item A.S. Rastorguev and A.K. Dambis, Astrophys. Bull. \textbf{66}, 47 (2011).

\item D. Reimers, \textit{Problems in stellar atmospheres and envelopes}
      (Ed. B. Baschek, W.H. Kegel, G. Traving, New York: Springer-Verlag, 1975), p. 229.

\item M.E. Sachkov,  Astron. Lett. \textbf{28}, 589 (2002).

\item D.G. Turner and L.N. Berdnikov, Astron. Astrophys. \textbf{423}, 335 (2004).

\item D.G. Turner and J.F. Burke, Astron. J. \textbf{124}, 2931 (2002).

\item B. Westerlund, Publ. Astron. Soc. Pacific \textbf{73}, 72 (1961).

\end{enumerate}

\newpage
\renewcommand{\thefootnote}{\arabic{footnote}}
\begin{threeparttable}
\caption{Fundamental parameters of long--period Cepheids}
\begin{center}
 \begin{tabular}{r|r|r|r|r|r|r|r|r}
  \hline
           &  $\Pi$ & $\mzams$ & $\tev$ &   $M$  &   $L$  &   $R$ & $\dot\Pi$ & $\dot\Pi_\star$ \\
           &  day   & $M_\odot$ & $10^6$~yr & $M_\odot$ &   $10^3L_\odot$  & $R_\odot$ & s/yr & s/yr \\
  \hline
 V1467 Cyg &  48.53 &    8.5 &   32.63 &   8.42 &   9.09 &   175 &   -170 &   -182\tnote{1} \\
    SV Vul &  45.06 &    8.8 &   30.51 &   8.71 &  10.91 &   182 &   -227 &   -214\tnote{2} \\
 V2641 Oph &  38.87 &    9.3 &   27.48 &   9.20 &  14.04 &   173 &   -320 &   -314\tnote{3} \\
    EV Aql &  33.27 &    9.3 &   27.48 &   9.20 &  14.21 &   158 &   -303 &   -310\tnote{1} \\
    RS Pup &  41.39 &   10.2 &   24.29 &  10.05 &  19.43 &   187 &     92 &    119\tnote{4} \\
    GY Sge &  51.54 &   10.7 &   22.31 &  10.53 &  22.23 &   214 &    192 &    203\tnote{5} \\
    II Car &  64.44 &   12.3 &   17.83 &  12.16 &  31.55 &   256 &    580 &    719\tnote{6} \\
  \hline
  \end{tabular}
  \begin{tablenotes}
  \item[1] Berdnikov (1994)
  \item[2] Turner, Berdnikov (2004)
  \item[3] Berdnikov et al. (2009a)
  \item[4] Berdnikov et al. (2009b)
  \item[5] Berdnikov et al. (2007)
  \item[6] Berdnikov, Turner (2010)
  \end{tablenotes}
\end{center}
\end{threeparttable}
\clearpage

\newpage
\clearpage

\newpage
\clearpage

\end{document}